\documentclass{article}
\usepackage{amscd,verbatim}
\usepackage[all]{xy}
\usepackage{graphicx}

\usepackage{amsmath}
\usepackage[psamsfonts]{amssymb}
\usepackage{amsthm}
\usepackage{euscript}

\usepackage{xypic}
\input{xypic}

\usepackage{epsfig}
\usepackage{amssymb,amsmath,amsthm,amscd}

\usepackage{latexsym}

\renewcommand{\(}{\begin{equation}}
\renewcommand{\)}{end{equation} \vspace{-.05in}\linebreak}

\newcounter{saveeqn}
\newcounter{savealpheqn}

\newcommand{\alpheqn}{\setcounter{saveeqn}{\value{equation}}%
  \stepcounter{saveeqn}\setcounter{equation}{0}%
  \renewcommand{\theequation}{\mbox{\arabic{section}.\arabic{saveeqn}
\alph{equation}}}
  \renewcommand{\)}{\end{equation}}}
\def\part#1{\frac{\partial}{\partial{#1}}}%
\def\group#1{\refstepcounter{equation}\setcounter{saveeqn}{\value{equati
on}}%
  \label{#1}\setcounter{equation}{0}%
\renewcommand{\theequation}{\mbox{\arabic{section}.\arabic{saveeqn}
\alph{equation}}}
  \renewcommand{\)}{\end{equation}}}
\newcommand{\reseteqn}{\setcounter{equation}{\value{saveeqn}}%
  \renewcommand{\theequation}{\arabic{section}.\arabic{equation}}%
  \renewcommand{\)}{\end{equation}}}

\newcommand{\aalpheqn}{\setcounter{saveeqn}{\value{equation}}%
  \stepcounter{saveeqn}\setcounter{equation}{0}%
  \renewcommand{\theequation}{\mbox{
        \Alph{subsection}.\arabic{saveeqn}\alph{equation}}}
   \renewcommand{\)}{\end{equation}}}
\newcommand{\areseteqn}{\setcounter{equation}{\value{saveeqn}}%
  \renewcommand{\theequation}{\Alph{subsection}.\arabic{equation}}%
  \renewcommand{\)}{\end{equation}}}

\renewcommand{\thefootnote}{\alph{footnote}}
\renewcommand{\(}{\begin{equation}}
\renewcommand{\)}{\end{equation}}
\newcommand{\ba}{\begin{eqnarray}}
\newcommand{\ea}{\end{eqnarray}}

\newcommand{\bp}{\mathop{\vtop{\ialign{##\crcr
   $\hfil\displaystyle{}\hfil$\crcr\noalign{\kern-13pt\nointerlineskip}
   \BIG{(}\hskip0pt\crcr\noalign{\kern3pt}}}}}
\newcommand{\cbp}{\mathop{\vtop{\ialign{##\crcr
   $\hfil\displaystyle{}\hfil$\crcr\noalign{\kern-13pt\nointerlineskip}
   \BIG{)}\hskip0pt\crcr\noalign{\kern3pt}}}}}
\newcommand{\pa}{\mathop{\vtop{\ialign{##\crcr

$\hfil\displaystyle{\oplus}\hfil$\crcr\noalign{\kern+1pt\nointerlineskip
}
   \hspace{.08in}$^{\alpha=0}$\hskip6pt\crcr\noalign{\kern3pt}}}}}

\newcommand{\R}{\ensuremath{\mathbb R}}

\newcommand{\cE}{\ensuremath{\mathcal E}}
\newcommand{\cF}{\ensuremath{\mathcal F}}

\newcommand{\C}{\ensuremath{\mathbb C}}

\newcommand{\Q}{\ensuremath{\mathbb Q}}

\newcommand{\Z}{\ensuremath{\mathbb Z}}

\def\O{\ensuremath{{\mathcal O}}}

\newcommand{\beq}{\begin{equation}}
\newcommand{\eeq}{\end{equation}}

\numberwithin{equation}{section}

\catcode`\@=11
\def\vereq#1#2{\lower3pt\vbox{\baselineskip1.5pt \lineskip1.5pt
\ialign{$\m@th#1\hfill##\hfil$\crcr#2\crcr\sim\crcr}}}
\catcode`\@=12

\makeatletter
\newcommand\figcaption{\def\@captype{figure}\caption}
\newcommand\tabcaption{\def\@captype{table}\caption}
\makeatother
\renewcommand{\(}{\begin{equation}}
\renewcommand{\)}{\end{equation}}

\newcommand{\bea}{\begin{eqnarray}}
\newcommand{\eea}{\end{eqnarray}}

\theoremstyle{plain}
\newtheorem{theorem}{Theorem}[section]
\newtheorem{lemma}[theorem]{Lemma}
\newtheorem{prop}[theorem]{Proposition}

\newtheorem{observation}[theorem]{Observation}

\theoremstyle{definition}
\newtheorem{definition}[theorem]{Definition}

\begin{document}

\begin{titlepage}
%\begin{flushright}

%hep-th/yymmxxx
%\end{flushright}

\vspace{2em}
\def\thefootnote{\fnsymbol{footnote}}

\begin{center}
{\Large\bf  Anomalies of $E_8$ gauge theory on String manifolds}
\end{center}
\vspace{1em}

\begin{center}
{\Large 
Hisham Sati
\footnote{Email:
{\tt hisham.sati@yale.edu}.
Current address: Department of Mathematics, University of Maryland, College Park, MD 20742.
Current Email: {\tt hsati@math.umd.edu}
}
}
\end{center}

\begin{center}
\vspace{1em}
{\em { Department of Mathematics\\
Yale University\\
New Haven, CT 06520\\
USA}}\\
\end{center}

\vspace{0.5cm}
\begin{abstract}
In this note we revisit the subject of anomaly cancelation in string theory and M-theory
on manifolds with String structure and give three observations. First, that on String manifolds 
there is no $E_8 \times E_8$ global anomaly in heterotic string theory. Second, 
that the description of the anomaly in the phase of the M-theory partition 
function of Diaconescu-Moore-Witten extends from the Spin case to the String case.
Third, that the cubic refinement law of Diaconescu-Freed-Moore for the phase of the
M-theory partition function extends to String manifolds.  
The analysis relies on extending from invariants which depend on the Spin structure
to invariants which instead depend on the String structure. 
Along the way, the one-loop term is refined via the Witten genus.

\end{abstract}

\tableofcontents

\vfill

\end{titlepage}
\setcounter{footnote}{0}
\renewcommand{\thefootnote}{\arabic{footnote}}

\pagebreak
\renewcommand{\thepage}{\arabic{page}}

%%%%%%%
%\section{Referee report}
%%%%%%%%

%
%The main claim of this paper is that the global anomaly of the heterotic
%$E8 \times E8$ and the anomaly in the M-theory partition function vanish on manifolds
%with String structure. This claim is based on certain results on String
%bordism groups proven in a separate paper. While those results are of mathematical
%interest, it is by no means clear that the applications discussed in
%the present work are physically sensible. 
%The problem is that the topological
%invariants in the heterotic or M-theory partition function depend explicitly
%on the choice of a Spin structure since Dirac operators are involved. That
%is why the question of anomaly cancellation involves Spin bordism groups.
%However they do not depend by construction on the choice of a String structure.
%Therefore, one simply does not need to consider extension problems for
%manifolds with String structure. This would make more sense for topological
%invariants which depend explicitly on the choice of a String structure. Then
%all questions raised in this paper would become relevant.
%I cannot recommend publication in the present form. It would be interesting
%if the author could construct a certain refinement of the M theory partition
%function relying on topological invariants which depend on the choice
%of a String structure. Then I think the paper would be very interesting and
%should be published.

%%%%%%%%%%%%%%%%%%%%%%%%%%%%%%%%
\section{Introduction}
%%%%%%%%%%%%%%%%%%%%%%%%%%%%%%%%

The DMW anomaly \cite{DMW} 
in the comparison of the partition functions of M-theory and
type IIA string theory is given by the vanishing of the seventh integral
Stiefel-Whitney class $W_7(X)$ of spacetime $X$. The cancelation 
of this anomaly in \cite{KS1} leads to the emergence of elliptic 
cohomology and other generalized cohomology theories. 
The ``String condition", i.e. the vanishing of half the Pontrjagin class
$\frac{1}{2}p_1(X)$ is stronger than vanishing of $W_7$. There are also
connections to generalized cohomology from the perspective of 
type IIB (target) string theory \cite{KS2} \cite{KS3}.  It is natural then
to ask how much of a role the String condition plays in global aspects of
string theory. It is the purpose of this note to provide a few observations that
provide one step in shedding some light on this question.  

\vspace{3mm}
Manifolds satisfying the String condition are called `` String manifolds" and are
characterized by having a String structure, i.e. 
with a lifting of the structure group of the tangent bundle from 
${\rm Spin}=O\langle 3 \rangle$ to its 3-connected cover 
${\rm String}= O \langle 7 \rangle$. Generic examples of String manifolds
occur when the manifold is highly connected. The 
simplest case is perhaps the spheres, which occur in the compactification
of eleven- and ten-dimensional supergravity (coupled to Yang-Mills)
leading to gauged supergravity in lower dimensions (see \cite{DNP}).
While compact manifolds $X$ with special holonomy $G_2$ and ${\rm Spin}(7)$
require a non-vanishing $p_1(X) \neq 0 \in H^4(X;\Z)$ \cite{J}, topological 
generalized such structures can be satisfied for manifolds with vanishing 
$p_1$ \cite{W}: $S^7$ admits a topological generalized $G_2$ structure
and any manifold $J$ of the form $S^1 \times N^7$, where $N^7$ is a Spin manifold, 
admits a topological generalized ${\rm Spin}(7)$ structure, since $J$ 
(trivially) satisfies $p_1(J)^2-4p_2(J)=0$ and $\chi(J)=0$. Note that 
there is a difference between $p_1$ being zero and $\frac{1}{2}p_1$ being 
zero, due to the possible existence of 2-torsion. Furthermore, there even exist {\it flat} 
manifolds of finite holonomy groups that have $p_1 \neq 0 
\in H^4\Z$ \cite{Z}. This is one justification for their use in models 
satisfying the equations of motion. 

\vspace{3mm}
The goal of this note is to point out the following:
\\
\noindent  {\bf (1)} The global anomaly for heterotic string theory vanishes on
String manifolds. 

\noindent {\bf (2)} The partition functions of M-theory and type IIA string theory 
match in the sense of \cite{DMW} on String manifolds. 

\noindent {\bf (3)} The description of the phase of the M-theory partition function
as a cubic refinement \cite{DFM} extends to String manifolds. 

\vspace{3mm}
One might wonder whether all what one needs to do to go from the
Spin case to the string case is set the first Pontrjagin class of the tangent 
bundle to zero, which would essentially make trivial the task we set out 
to achieve. This turns out to be naive because we are considering 
global questions. In particular, we are not guaranteed that the 
desired obstructions vanish. More precisely, we are trying to extend 
bundles on String manifolds rather that on Spin manifolds, and the 
corresponding cobordism groups may introduce new invariants and
obstructions in going from the Spin to the String case. That
we show that this is not the case is not immediate and is in fact nontrivial.
Mathematically, the results are possible due to the recent calculations 
of $M{\rm String}_n\left( K(\Z, 4)\right)$ for $n\leq 14$ in 
the companion paper \cite{Hill}. That paper provides the main 
technical topological results and this note provides the 
analysis and geometry as well as 
physical motivation and the corresponding consequences. 

\vspace{3mm}
 One might raise the following objection: If the anomaly vanishes for 
weaker condition then should it not vanish for the stronger one? The answer
is not obvious as all without nontrivial calculations, and in fact the notions
of  ``weaker" and ``stronger" might be misleading in this context.
We know that 
$\Omega_{11}^{\rm spin}\left(K(\Z, 4)\right)=0$ but we are not guaranteed
that the corresponding cobordism group on String manifolds vanish. We show
that this is the case, i.e. that $M{\rm String}_{11}\left(K(\Z, 4)\right)$ is zero,
and hence there are no new obstructions and the extension continues to 
be possible.  
A less nontrivial point is to check whether the string condition is preserved in 
forming the 11-dimensional mapping torus and its corresponding 
12-dimensional bounding manifold. The global anomaly vanishes
for $S^{10}$ \cite{CMP}. 

\vspace{3mm}
The objection would likewise be on both the cancelation of the DMW 
anomaly and the matching of the phases of the partition function. On the former
there are indeed no subtleties, but the latter requires the analysis of subtle
mod 2 invariants. In particular the question of extension of the $\Z_2$ invariant
$f(a+b)- f(a) -f(b)$ of \cite{DMW} becomes a question of extension on String
manifolds of $M{\rm String}_{10} ( Z)$ rather than $M{\rm Spin}(Z)$ for the Spin
case, where $Z$ is $K(\Z, 4 ) \times K(\Z, 4)$ or $K(\Z, 4) \wedge K(\Z, 4)$. 
Indeed one of the results is that  $M{\rm String}_{10} ( Z)= M{\rm Spin}_{10}(Z)$,
so that there are no new invariants, hence obstructions, beyond the one 
coming from the Spin case.

\paragraph{Spin vs. String invariants.}
The topological
invariants in the heterotic or the M-theory partition functions depend explicitly
on the choice of a Spin structure since Dirac operators are involved. That
is why the question of anomaly cancellation involves Spin bordism groups.
However they do not depend by construction on the choice of a String structure.
Therefore, for those original invariants, 
one simply does not need to consider extension problems for
manifolds with String structure.  However, this would 
be needed for topological
invariants which depend explicitly on the choice of a String structure.
To that end, we discuss a certain refinement of the M theory partition
function relying on topological invariants which depend on the choice
of a String structure. 

%\attn{Move from here}

 %\item
 \vspace{3mm}
% In \cite{IV} one of the 
% two topological terms in the M-theory
% action, namely 
 We study the one-loop term in M-theory and 
 provide a refinement to $q$-expansions via the 
 Witten genus.  
 The consequences for M-theory are as follows.
 The one-loop polynomial 
 $I_8$ appearing from the M5-brane worldvolume 
 corresponds in the M-theory Lagrangian, 
 via anomaly inflow, to the one-loop term $I_{\rm 1-loop}=-I_8\wedge C_3$
\cite{VW}  \cite{DLM}.
 Thus, this topological term in M-theory will admit a 
 similar elliptic refinement. As the topological terms in 
 M-theory, namely the Chern-Simons term $I_{CS}=\frac{1}{3}C_3 \wedge G_4 \wedge G_4$
 and the one-loop
 term
 $I_{\rm 1-loop}$, together with the Rarita-Schwinger field, make up 
 the main contribution to the partition function of the C-field,
 this leads to $q$-expansions in the M-theory action.
 We provide another interpretation of such expansions 
 via replacing ordinary $G$ bundles with corresponding 
 loop bundles. 
 The phase of the partition function will also admit 
 such a refinement. We study this via a refinement of the 
 Atiyah-Patodi-Singer index theorem \cite{APS} to the 
 case of $q$-bundles in the sense of \cite{BN} \cite{Ga}. 
 The dimensional reduction, via the 
 adiabatic limit on the M-theory circle, of the 
 elliptically refined eta invariant will lead to 
 refined 
 eta-forms in ten dimensions, making connection to the discussion
 in \cite{S-gerbe}. In type IIA, the $q$-refinement of the mod 2 index of the
 Dirac operator in relation to the partition function has already been 
 discussed in \cite{KS1} \cite{S-op}.

% By anomaly inflow, the one-loop term $I_8$

% \attn{All corrections Green are modular forms. Has to do with the membrane
% worldvolume}
 
 \vspace{3mm}
 Eleven-dimensional supersymmetry relates the one-loop term 
$C_3 \wedge I_8$ to a (curvature)${}^4$-term \cite{GV}
and to other (yet not fully 
determined) terms that enter in the effective action. 
While the effective action of the compactified theory depends nontrivially
on moduli fields associated with the geometry of the compact manifold,
this dependence is restricted by consistency with the duality symmetries
in ten and lower dimensions. For tori, this demonstrated in \cite{GGV}
\cite{GRV}.
The exact BPS (curvature)${}^4$-couplings in M-theory compactified on a 
torus are obtained in \cite{PNPW} \cite{PW}
 from the toroidal BPS membrane,
 using the point of view that the eleven-dimensional membrane gives the fundamental degrees of freedom of M-theory. This involves modularity coming from wrapped membranes. 

%\vspace{3mm} 
%The one-loop term $I_8$ was refin
%the one-loop {\it polynomial}. This is related directly to the worldvolume
%anomaly of the M5-brane. We call $C_3 \wedge I_8$ the corresponding 
%one-loop {\it term}. This is the term in the eleven-dimensional action that leads to the
%cancellation of the anomaly of the M5-brane. 

%\attn{until here}

 \vspace{3mm}
 In section \ref{Sec refine} we provide the invariants which depend on the String structure.
 Later in section \ref{Sec results} we give the results listed above in the third paragraph
 of the introduction

 %%%%%%%%%%%%%%
\section{The Refined Invariants}
%%%%%%%%%%%%%%
\label{Sec refine}
In this section, we propose the invariants that would replace those invariants which
depend on the choice os Spin structure. The invariants we use will likewise depend
(in general) on the choice of String structure. Explicit examples are given in 
\cite{tcu}, following the constructions of \cite{Red} \cite{Bu} \cite{BN}.
This can be seen most easily for the case of a three-dimensional sphere where the
choice of framing is the same as the choice of String structure, on which the invariants
depend. In our current case we can, for instance, take our eleven-manifolds $Y^{11}$ 
to be products of $S^3$ with eight-dimensional factors. 

%%%%%%%%%%%%%%%%%%%%
\subsection{Motivation for the invariants}
%%%%%%%%%%%%%%%%
%\section{Analyzing the one-loop term} 
 %%%%%%%%%%%%
 
 We start by providing motivation for refinement of the one-loop term
in M-theory.
% analogously to what we did for the one-loop polynomial in the
%context of M5-branes in \cite{IV}.

 \paragraph{$I_8$ vs. Index and why String structure: from type IIA.}
 Consider the topological part of the action in type IIA string theory,
 which involves the $B$-field with its field strength $H_3$.
 Since $I_8$ does not have integral periods then a coupling of the form
 $\exp [2\pi i \int_{X^{10}} B_2 \wedge I_8]$ would not be well-defined in 
 the presence of topologically nontrivial $B$-fields. This led the authors of 
 \cite{BM} 
 to provide the following solution. The last sentence in the previous paragraph
  means that the term 
 $\exp [2\pi i \int_{X^{10}} B_2 \wedge 30\widehat{A}_8]$ is well-defined. 
 However, in order for the remaining term 
 $\exp [2\pi i \int_{X^{10}} B_2\wedge \frac{1}{8}\lambda^2]$ to be obtained 
 from a gauge-invariant partition function, the condition 
 $[(\lambda+ 2\rho) \wedge H_3]_{\rm dR}=0$
  is proposed in \cite{BM}, 
 where $\rho$ is a closed 4-form with integral periods. 
 Now we see that one simple way of satisfying the above condition is 
to require triviality of $\lambda$. This means that we have a String 
structure, that is our manifold $X^8$ has a lifting of the structure group 
to the 3-connected cover of the Spin$(10)$. 
 So what we have 
 is an expression, which is 
essentially an index, valid when we impose the String condition.
 This observation will be important for us.
 
% \vspace{3mm}
 %To warm up to the case of type IIA, let us consider heterotic string theory.

 \paragraph{Anomaly cancellation via the elliptic genus:
 from heterotic string theory.} 
% from the 
 %elliptic genus.}
 Consider heterotic string theory on a Spin ten-manifold $M^{10}$. 
Here in addition to the tangent bundle there is also a gauge 
bundle with connection of curvature $F$.
 Because of modular invariance, the anomaly in this theory
 always has a factorized form \cite{SWa}
 \(
 I_{12}(F,R)=\frac{1}{2\pi}({\rm Tr}F^2 - {\rm Tr}R^2) \wedge X_8(F,R)\;,
 \)
 where $X_8(F,R)$ is the Green-Schwarz anomaly polynomial. 
 This anomaly can be cancelled by the following term in the effective 
 action \cite{GS} \cite{GSW} \cite{LNS}
 \(
 S_c= \int_{M^{10}} B_2 \wedge X_8(F,R)\;.
 \)
 There is an interpretation in terms of the elliptic genus \cite{LNS} \cite{LNSW}.
  The anomaly is 
 given by the constant term of weight two of the 12-form in the 
elliptic genus
 \(
 I_{12}(F,R)= \phi_{\rm ell}(q, F, R)\vert{{}_{12-{\rm form~ coeff.~ of~} q^0}}\;.
 \)
 The terms which violate modular invariance
 can be factored out of the character-valued partition function
 \(
 \phi_{\rm ell}(q, F, R)= \exp \left( -\frac{1}{64\pi^4}E_2(q) 
 ({\rm Tr}F^2 - {\rm Tr}R^2)\right) \widetilde{\phi}_{\rm ell}(q, F, R)\;,
 \)
where $\widetilde{\phi}_{\rm ell}(q, F, R)$ is fully holomorphic 
and modular invariant of weight 4, that is it can be expressed in terms of 
the Eisenstein function $E_4(q)$, and $E_2(q)$ is the 
non-modular Eisenstein series of weight 2 (see expression \eqref{e2}).
These can be written in terms of a parameter $\tau \in \mathcal{H}$
in the complex upper half plane via $q=e^{2\pi i \tau }$.
 Interestingly, the anomaly canceling terms are encoded in 
 a $\tau$-integral over the fundamental domain $\mathcal{F}$
of the weight zero terms of the modified elliptic genus
\cite{LNSW}
 \(
 \overline{\phi}_{\rm ell}(q, F, R)=
 \exp \left( -\frac{1}{64\pi^4}\frac{\pi}{ {\rm Im} \tau} 
 ({\rm Tr}F^2 - {\rm Tr}R^2)\right) \widetilde{\phi}_{\rm ell}(q, F, R)\;,
 \)
namely 
\(
-\frac{1}{64\pi^2} \int_{\cF} \frac{d^2\tau}{({\rm Im}\tau)^2}
\overline{\phi}_{\rm ell}(q, F, R)|_{\rm 8-forms}= X_8(F, R)\;.
\)
For comparison with our case, let us highlight some features of this \cite{LNS} \cite{LNSW}:
 \begin{enumerate}
 \item Essentially the same function that captures the anomaly also
 governs the cancellation of that anomaly. 
 
 \item The integrands have a nontrivial dependence on the modular parameter
 $\tau$, reflecting contributions not only from physical massless states 
 but also from an infinity of `unphysical' modes. This is in contrast to 
 what happens in field theory, signaling that string theory should not be thought
 of merely as a superposition of an infinite number of point particle 
 field theories.

 \item The interest in the structure of the
 anomaly canceling terms was not to prove 
  {\it that} the anomaly cancels, but to understand {\it how}
  it cancels.  
 \end{enumerate}
 
 Now in the case of M-theory, we can have essentially the same line of reasoning 
 for all three points above, which refer to heterotic string theory. 
 First, the anomaly involves $I_8$, which is the same term (up to a sign)
 which cancels that anomaly. Second,
 M-theory is more unlike field theory than is string theory, and hence
 we also expect an important role for seemingly unphysical 
 infinite modes. Third, indeed, understanding how the anomaly 
 cancels allows us to provide a generalization, namely the 
 elliptic refinement.  
 We will explore similar features for type IIA string theory. 
  The case for similarity between heterotic string theory and 
  type IIA string theory is already made in \cite{VW}, which we 
  consider next to further motivate our proposal.
   
 \paragraph{The one-loop term in type IIA via modularity.}
Duality, such as that between heterotic string theory and type IIA string theory,
 tends to exchange worldsheet and spacetime 
 effects (cf. \cite{DL}). 
 Torsion in cohomology shows up in the worldsheet global anomalies
in the heterotic string, so that on the type IIA side this would 
appear in {\it spacetime} global anomalies.  
So we expect a similar situation as in \cite{LNSW} to take
place on the spacetime side in type IIA. Let us highlight  an important 
aspect of this.  
 In \cite{VW} it is shown that in ten-dimensional type IIA there is a one-loop 
 contribution to the effective action of the form 
 $
 \delta S= \int_{X^{10}} B_2 \wedge I_8$.
 %\;.
 %\label{VW term}
 %\)
 The computations that lead to
 this term
%  \eqref{VW term} 
are 
 similar to those of \cite{LNSW} leading to the above anomaly 
 cancellation in heterotic string theory, even though 
 type IIA string theory in non-chiral. 
 The one-loop computation on $X^{10}=T^2 \times M^8$
  involves integrating over the 
 fundamental domain $\mathcal{M}$
 corresponding to odd (even) Spin structure 
 for the left- (right-) moving fermions on the torus. 
 The integral is 
 \(
 \delta S= B_2\cdot \frac{-i}{4}\int_{\mathcal{M}} \frac{d^2\tau}{2\pi {\rm Im} \tau}
 \cdot \frac{\pi}{{\rm Im} \tau} \phi_{\rm ell}(q)\;.
 \)
 This is reduced to an integral over the boundary of the 
 moduli space. Furthermore, the elliptic genus is replaced with 
 the massless contribution:
\begin{enumerate}
\item  In the (NS, R) sector, which is fermionic, the massless contribution of $\phi_{\rm ell}(q)$
 computes the index $n^{NS,R}$ of the Dirac operator coupled to $TM$ given by 
 % anomaly polynomials
$n^{NS, R}:= \int_{M^8} I_8^{NS, R}= \int_{M^8} 2 I_8^B + 3I_8^\lambda$.
 
\item In the (R, R) sector, which is bosonic,
 the massless contribution of $\phi_{\rm ell}(q)$
 computes the index $n^{R,R}$ of the Dirac operator coupled to the Spin 
 bundle $SM$
 given by $n^{R,R} := \int_{M^8} I_8^{R,R}= \int_{M^8} -2I_8^B$. 
 \end{enumerate}
\noindent  Then the result in two dimensions is 
 $
 \delta S= \frac{i\pi}{24} (2n^{NS, R} - n^{R,R}) B_2
 $. Thus, the one-loop polynomial (and hence one loop term) in type IIA
 enters in the anomaly via the elliptic genus.
 % where the indices are integrals of anomaly polynomials of the form 
% \bea
% n^{NS, R}&:=& \int_{M^8} I_8^{NS, R}= \int_{M^8} 2 I_8^B + 3I_8^\lambda 
% \nonumber\\
% n^{R,R} &:=& \int_{M^8} I_8^{R,R}= \int_{M^8} -2I_8^B
% \eea

%\attn{ so that }

 %\end{enumerate}

 \paragraph{Type IIB string theory.}
 %\begin{enumerate}
  This string theory is invariant under worldsheet parity. However, 
 the term $\int B_2 \wedge I_8$ is odd under $B_2 \mapsto -B_2$, so that 
such a term is absent in type IIB string theory.
 Indeed, this is demonstrated explicitly in \cite{DM}.

\paragraph{Topological String structures.}
A topological String structure $\alpha^{\rm top}$ is by definition 
a ``trivialization" of the Spin characteristic class 
$\frac{1}{2}p_1(TM) \in H^4(M;\Z)$ (see \cite{Bu} \cite{BN}).
A String manifold $(M, \alpha^{\rm top})$ is a Spin manifold 
$M$ with a chosen topological String structure $\alpha^{\rm top}$ on
the Spin bundle. 
The Thom spectrum MString can be realized as bordism classes of
closed String manifolds $[M,\alpha^{\rm top}]$, i.e. manifolds with
a String structure on their tangent bundle. The morphism
$i: {\rm MString}_* \to {\rm MSpin}_*$ forgets the topological 
String structure, i.e. is given 
%geometrically 
by
$i([M, \alpha^{\rm top}])=[M]$.

 %\paragraph{Kriz-Xing.} 
 % \vspace{3mm}
  \paragraph{Remark on connection to other proposals.}
  A proposal for refining
  field strength $G_4$  of the C-field is given in \cite{KX}, where the main idea is to 
  replace $G_4$ essentially with $G_4 E_4(q)$, where $E_4(q)$ is the 
  Eisentein series, a modular form of weight 4. 
  %Although this interesting replacement
  %s somewhat ad hoc, it 
  This replacement 
  is related
  to what we propose in this
  paper in the following sense. Consider the one-loop term 
 lifted to twelve dimensions, that is $G_4 \wedge I_8$. This terms
 gets $q$-expansions via refining either $G_4$, or $I_8$ (or both).
 What we do in this paper is the latter. Since the corresponding 
 refinement involves $E_4(q)$, we see that both refinements
 are essentially the same, as far as the one-loop term is concerned. 
% end result within M-theory. 
 However, the method we provide is 
 motivated by anomaly cancellations and hence seems more 
 physically justified.

 \subsection{The Witten genus}

% \subsection{The Witten genus}
%\vspace{3mm}
The Witten genus is given by the series in the formal variables $x$ 
and $q$ (see \cite{HBJ})
\bea
\phi_W (x, q)&:=& \frac{\frac{x}{2}}{{\rm sinh} (\frac{x}{2})} \prod_{n=1}^\infty 
\frac{(1-q^n)^2}{(1-q^ne^x)(1-q^n e^{-x})}
\nonumber\\
&=&
\exp \left[ \sum_{k=2}^\infty \frac{2}{(2k)!}E_{2k}(q)x^{2k}\right]
e^{E_2 (q) x^2}
\nonumber\\
&=& \theta_W (x, q) e^{E_2 (q) x^2} \in  \Q [[q]][[x]]\;.
\eea
%with  $\theta_W (x, q) \in  \Q [[q]][[x]]$.
Here
$
E_{2k}(q):= -\frac{B_{2k}}{4k} + \sum_{n=1}^\infty \sigma_{2k-1}(n)q^n \in \Q [[q]]
$ is the Eisenstein series,
with the Bernoulli numbers $B_{2k}$ and the sum $\sigma_{2k-1}(n)$ of the 
$(2k-1)$th powers of the positive divisors of $n$. For $k \geq 2$ the power series 
$E_{2k}(q)$ is the 
$q$-expansion of a modular form of weight $2k$.
The constant term in $E_{2k}(q)$ is rational 
while the higher terms are integral.  
% \paragraph{The non-modular Einsenstein function $E_2(q)$.}
 The Einsenstein function $E_2=\sum_{m,n\in \Z} (m\tau +n)^{-2}$ is 
 not a modular function, but transforms in an anomalous way
 \(
 E_2\left( \frac{a\tau + b}{c\tau +d}\right)=(c\tau + d)^2 E_2(\tau) - 2\pi i c (c\tau + d)\;.
 \label{e2}
 \)
 A modular function $\hat{E}_2(\tau)$ can be built out of $E_2(\tau)$
on the expense of losing holomorphicity in $\tau$, 
$
\hat{E}_2(\tau)= E_2(\tau) - {\pi}/{{\rm Im}(\tau)}$.
Note that $\phi_W (x, 0)$ is the $\widehat{A}$-genus, with 
the formal variable $x$ giving the Pontrjagin classes via the 
generating formula $\sum_{i\geq 0}p_i= \prod_{i=1}^\infty (1+ x_i^2)$. 
When $x^2=0$, i.e. $p_1=0$, the dependence of the Witten genus on 
$E_2$ is removed and hence modularity is restored \cite{Za}.

\paragraph{Multiplicative series and $q$-expansions.}
%Thus, define 
%\(
%\theta_W (x,q):= \exp \left[ 
%\sum_{k=2}^\infty \frac{2}{(2k)!}E_{2k}x^{2k}
%\right] \in \Q [[q]][[x]]\;.
%\)
For $k\geq 1$, consider the Newton polynomials
$
N_{2k}(p_1, \cdots)=\sum_{j=1}^\infty x_j^{2k} \in \Q [[p_1, p_2, \cdots]]
$,
and consider the two series in $\Q [[q]][[p_1, p_2, \cdots ]]$ defined in 
\cite{BN} as
\begin{eqnarray}
\Phi (p_1, \cdots) &=& \exp \left[ 
\sum_{k=2}^\infty \frac{2}{(2k)!}E_{2k}N_{2k}(p_1, \cdots)
\right]e^{E_2p_1} 
\label{fi}
\\
\widetilde{\Phi}(p_1, \cdots) 
&=& \exp \left[ 
\sum_{k=2}^\infty \frac{2}{(2k)!}E_{2k}N_{2k}(p_1, \cdots)
\right]\sum_{j=1}^\infty \frac{E_2^jp_1^{j-1}}{j!}=
 \Theta \frac{e^{E_2 p_1}-1}{p_1}\;,
\label{fi t}
\end{eqnarray}
where $\Theta:=\prod_{i=1}^\infty \theta_W(x_i) 
\in \Q[[q]][[p_1, p_2, \cdots]]$.
With $N_2=\sum_{j=1}^\infty x_j^2=p_1$
and $N_4=\sum_{j=1}^\infty x_j^4=p_1^2-2p_2$,
we find that 
the degree eight components of \eqref{fi} and \eqref{fi t}, respectively,
 are
\begin{eqnarray}
\Phi_8 &=& \frac{1}{12}E_4 (p_1^2- 2p_2) + \frac{1}{2} E_2^2p_1^2\;,
\\
\widetilde{\Phi}_8 &=& 
E_2 \left[ \frac{1}{12}E_4 (p_1^2- 2p_2) + \frac{1}{3!} E_2^2p_1^2
\right]\;.
\end{eqnarray}
Imposing the String condition, $\frac{1}{2}p_1=0$ (or, rationally,
$p_1=0$), gives
\(
\Phi_8= -\frac{1}{6}E_4 p_2\;, \qquad \qquad
\widetilde{\Phi}_8= -\frac{1}{6}E_2 E_4 p_2\;.
\label{q phi}
\)
%\paragraph{The $q$-expansions.}
Using 
$E_4= 1+ 240 \sum_{n=1}^\infty \frac{n^3 q^{2n}}{1-q^{2n}}$
%\attn{$N^3$ dofs?????}
and 
$E_2=-\frac{1}{24} + q + \cdots$, 
the expressions \eqref{q phi} admit the following $q$-expansions 
\(
\Phi_8=-\frac{1}{6}p_2 + \O(q)\;,
\qquad \qquad
\widetilde{\Phi}_8=\frac{1}{144}p_2 + \O(q)\;.
\)
We therefore see that the $q=0$ term of 
$\Phi_8$ is, up to a sign, the obstruction to having a
Fivebrane structure \cite{SSS2} \cite{SSS3} and 
the $q=0$ term of $\widetilde{\Phi}_8$ is three times the one-loop term 
$I_8$.

\paragraph{Virtual bundles.}
The above genera have another description, namely, one
can view them as corresponding to ordinary Dirac operators
coupled to infinite formal symmetric powers of an underlying
finite-dimensional bundle. 
Corresponding to a real $l$-dimensional bundle $V$, define the formal power series
of virtual bundles 
\(
{\sf R}(V):=\sum_{n\geq 0} q^n {\sf R}_n(V) =\prod_{k\geq 1}
(1-q)^{2l} 
\bigotimes_{k\geq 1} {\rm Sym}_{q^k}(V \otimes_\R \C)\;, 
\label{virt}
\) 
where ${\rm Sym}_q(V):=\bigoplus_{n\geq 0} q^n {\rm Sym}^n(V)$ is the generating 
power series of the symmetric powers ${\rm Sym}^n(V)$ of the bundle $V$.
For a manifold $M$, 
the following topological index formula holds \cite{BN}
\(
\Phi(p_1(TM), \cdots)= \widehat{A} (TM) \cup {\rm ch}({\sf R}(TM))\;.
\label{ph 12}
\)

%%%%%%%%%%%%%%%%
%\section{Consequences for M-theory}
%%%%%%%%%%%%

\subsection{Refined geometric invariants for the topological terms in M-theory}
%As in the case of the M5-brane, a 

\paragraph{Geometric String structures.}
A geometric refinement $\alpha$ of  the topological String structure
$\alpha^{\rm top}$
trivializes this class at the level of differential forms, 
that is \cite{Red} \cite{Wal} $\alpha$ gives rise to a form 
$C_\alpha \in \Omega^3(Y^{11})$ such that 
$dC_\alpha =\frac{1}{2}p_1(\nabla^m)$.
This is essentially the C-field as explained in \cite{tcu}.
If $\nabla^V$ is a connection on a real vector bundle 
$V \to Y^{11}$ over some manifold $Y^{11}$, then the Chern-Weil 
representative of the $i$th Pontrjagin 
class will be denoted $p_i(\nabla^V) \in H^{4i}(Y^{11};\Z)$.
Let $Y^{11}$ be a closed seven-manifold with a Spin structure $SY^{11}$ and 
a topological String structure $\alpha^{\rm top}$ and 
Levi-Civita connection $\nabla^{TY}$. 
Let $V \to Y^{11}$ be a real vector bundle with a metric 
$h^V$ and connection $\nabla^V$. The twisted Dirac operator $D_Y \otimes V$ 
acts on sections of the bundle $SY^{11} \otimes_\R V$. 
Call a Riemannian Spin manifold a {\it geometric manifold} $\mathcal{Y}$. In the presence of 
$V$, we also have a {\it geometric bundle} $\mathcal{V}:=(V, h^V, \nabla^V)$, in 
the sense of \cite{BN}. 
%\paragraph{Taming.} 
The refined eta invariants require a taming \cite{BN}.
A {\it taming} of $D_Y \otimes V$ is a self-adjoint operator 
$Q$ acting on $\Gamma(S(Y^{11}) \otimes V)$ and given by a smooth 
integral kernel such that $D_Y \otimes V+Q$ is invertible. 
Such a taming always exists in eleven dimensions.
In Physical terms, we interpret a taming as the Pauli-Villars regularization 
imaginary mass term $im$.

%\vspace{3mm}
%A geometric refinement $\alpha$ of  the topological String structure
%$\alpha^{\rm top}$
% gives rise to a form 
%$C_\alpha \in \Omega^3(Y^{11})$ such that 
%$dC_\alpha =\frac{1}{2}p_1(\nabla^Y)$.
%This is essentially the C-field as explained in \cite{tcu}.
%If $\nabla^V$ is a connection on a real vector bundle 
%$V \to M$ over some manifold $M$, then the Chern-Weil 
%representative of the $i$th Pontrjagin 
%class will be denoted $p_i(V) \in H^{4i}(M;\Z)$.
%Let $Y^{11}$ be a closed 11-manifold with a Spin bundle $SY^{11}$, 
%a topological String structure $\alpha^{\rm top}$ and 
%Levi-Civita connection $\nabla^{TY}$. The Riemannian metric on $Y^{11}$ gives a Levi-Civita connection $\nabla^Y$. 
%Let $V \to Y^{11}$ be a real vector bundle (which we think of as the $E_8$ bundle)
%with a metric 
%$h^V$ and connection $\nabla^V$. The twisted Dirac operator $D_Y \otimes V$ 
%acts on sections of the bundle $SY^{11} \otimes_\R V$. 
%Call a Riemannian Spin manifold a {\it geometric manifold} $\mathcal{Y}$. In the presence of 
%$V$, we also have a {\it geometric bundle} $\mathcal{V}:=(V, h^V, \nabla^V)$, in 
%the sense of \cite{BN}. 
%%\paragraph{Taming.} 
%The refined eta invariants require a taming \cite{BN}.
%A {\it taming} of $D_Y \otimes V$ is a self-adjoint operator 
%$Q$ acting on $\Gamma(S(Y^{11}) \otimes V)$ and given by a smooth 
%integral kernel such that $D_Y \otimes V+Q$ is invertible. 
%Such a taming always exists in eleven dimensions.
%In Physical terms, a taming is the Pauli-Villars regularization 
%imaginary mass term $im$.

\paragraph{String structures 
and higher powers 
on $Z^{12}$ vs. on $Y^{11}$.}
Consider the String manifold $(Z^{12}, \beta^{\rm top})$ 
with topological String structure $\beta^{\rm top}$ and 
boundary $Y^{11}$. The tangent bundle decomposes as
$TZ^{12}|_Y \cong TY^{11} \oplus \O_\R$. The trivial bundle
$\O_\R:=Y^{11}\times \R$ has a canonical String structure. 
Then the decomposition implies that 
$Y^{11}$ has an induced 
String structure $\alpha^{\rm top}$. 
Thus, starting from a String twelve-manifold we get an eleven-dimensional 
String boundary with an induced topological String structure.
Let $[Z^{12}]\in {\rm MSpin}_{12}$ be a homotopy class represented by a closed 
12-dimensional Spin manifold $Z^{12}$. 
The Witten genus in twelve dimensions is a morphism of ring spectra 
$
{\sf R}: {\rm MSpin}_{12} \to KO[[q]]_{12}\cong\Z[[q]]$,
and is given by (see \cite{HBJ})
\(
{\sf R}([Z^{12}])=\frac{1}{2} {\rm Ind}(D_Z \otimes {\sf R}(TZ^{12}))
=\frac{1}{2} \langle \Phi, [Z^{12}] \rangle\;.
\)
%with ${\sf R}$ as in 

\paragraph{The secondary invariants.}
%\vspace{3mm}
%We now consider the bounding String twelve-manifold 
%$Z^{12}$. Choose a connection $\nabla^{TZ}$ on the tangent bundle 
%$TZ^{12}$ of $Z^{12}$. 
The secondary invariant $s(Y^{11}, \alpha)$ requires the existence 
of a Spin zero bordism $Z^{12}$
of the String manifold $(Y^{11}, \alpha)$. 
This is always possible as the String cobordism group in eleven 
dimensions is zero. 
%(see \cite{e8} for further discussions on this).
Choose a connection $\nabla^Z$ extending the connection $\nabla^Y$
on the tangent bundle $TY^{11}$ of $Y^{11}$. 
Consider also the trivial bundle $\O_\R=Y^{11}\times \R$ with the 
trivial connection. 
%where $\mathcal{O}_\R=Y^{11}\times \R$ is a trivial bundle 
%with a trivial connection 
resulting 
from the restriction $TZ^{12}|_{Y^{11}}\cong TY^{11} \oplus \mathcal{O}_\R$. 
Both connections induce a connection on the virtual 
bundles ${\sf R}_n(TY^{11} \oplus \O_\R)$ for all $n\geq 0$. 
%Choosing a geometric refinement of the topological 
%String structure gives a 3-form $C_\alpha \in \Omega^3(Y^{11})$ 
%satisfying $dC_\alpha=\frac{1}{2}p_1(\nabla^Y)$.
At the level of differential forms, for a twelve-manifold $Z^{12}$, 
formula \eqref{ph 12} gives
\(
\Phi (p_1(TZ), \cdots )=\widehat{A}(TZ^{12}) \cup {\rm ch} 
({\sf R}(TZ^{12}) \in \Omega^{4*}(Z^{12})[[q]]\;.
\)
Applying \cite{BN} to our case, the analytical and geometric secondary invariants are 
given by the formal power series
% \cite{BN}
\begin{eqnarray}
s^{\rm an}(Y^{11}, \alpha, t)&:=&
\int_{Y^{11}} C_\alpha \wedge
\widetilde{\Phi}_8(\nabla^Y) +\frac{1}{2} \sum_{n\geq 0} q^n 
\eta ({\sf R}_n (TY^{11}\oplus \O_\R)) \in \R[[q]]\;,
\\
s^{\rm geom}(Y^{11}, \alpha, \nabla^Z) &:=&
\int_{Y^{11}} C_\alpha \wedge
\widetilde{\Phi}_8(\nabla^Y)
- \frac{1}{2} \int_{Z^{12}} \Phi_{12} (\nabla^Z) \in \R[[q]]\;,
\label{sgeom}
\end{eqnarray}
where the expression $s^{\rm an}$ is a String cobordism class of the String manifold 
$(Y^{11}, \alpha^{\rm top})$, and 
the cohomology classes match $[s^{\rm an}] =[s^{\rm geom}]$.
Now, the integrands in \eqref{sgeom} involves the expressions
\begin{eqnarray}
\widetilde{\Phi}_8&=& \frac{1}{12}E_2 \left[
-E_4 p_2 + (2E_2^2 + E_4) p_1^2
\right]
\\
\Phi_{12}&=&\frac{1}{12}\left[ 
E_2 E_4 p_1(p_1^2-2p_2) + \frac{1}{20} E_6 (p_1^3-3p_1p_2+3p_3)
\right]\;.
\end{eqnarray}
Applying the String condition, these give 
\(
\widetilde{\Phi}_8|_{\lambda=0}=-\frac{1}{12}E_2 E_4 p_2=\frac{1}{2^5\cdot 3^2}
p_2+ \mathcal{O}(q) {\rm ~~~and~~~}
\Phi_{12}|_{\lambda=0}=\frac{1}{80}E_6 p_3=\frac{1}{2^3\cdot 3^2}p_3 + 
\mathcal{O}(q)\;,
\)
respectively.
Note that in the zero mode case in twelve dimensions,
the addition of the Chern-Simons
term $L_{CS}=\frac{1}{6}G_4\cup G_4 \cup G_4$ to the 
one-loop term $L_{1-{\rm loop}}=-I_8\wedge G_4$
leads to a cancellation of the top Pontrjagin class $p_3$.
The same would happen here so that the two expressions
coincide after taking cohomology classes, as in the case of the 
M5-brane. We see that the integrand in $s^{\rm an}(Y^{11}, \alpha, t)$ 
is of the form $\frac{1}{6}\int_{Y^{11}} C_\alpha \wedge I_8|_{\lambda=0}
+ \mathcal{O}(q)$.

%
%
%\vspace{3mm}
%\(
%s^{\rm an}=\int_{Y^{11}} C_\alpha \wedge \widetilde{\Phi}(\nabla^{TY}) 
%+ \frac{1}{2}\sum_{n\geq 0}
%q^n \eta \left(R_n(TY^{11} \oplus \mathcal{O}_\R) \right) \in \R[[q]]\;,
%\)
%
%
%\vspace{3mm}
%The formal power series 
%\(
%s^{\rm geom}(Y^{11}, \alpha, \nabla^{TZ})=\int_{Y^{11}} C_\alpha \wedge \widetilde{\Phi}(\nabla^{TY})
%-\frac{1}{2} \Phi (\nabla^{TZ}) \in \R[[q]]\;, 
%\)
%so that 
%\(
%\left[ s^{\rm geom}(Y^{11}, \alpha, \nabla^{TZ})\right]=
%s^{\rm an} (Y^{11}, \alpha^{\rm top})\;.
%\)
%

%\attn{KS1 $q=v_1^3 v_2^{-1}$}

\subsection{The reduction to ten dimensions and eta forms}

We now consider $Y^{11}$ to be a circle bundle 
 over a ten-manifold $X^{10}$, over which type IIA string theory is taken. We consider the 
 effects of the modes of this M-theory circle on analytical quantities which enter
 into the definition of the phase of the partition function.

 \paragraph{Higher powers as higher modes of the 11th Rarita-Schwinger bundle.}
  Let $S(S^1)$ be the Spin bundle associated to the vertical 
 tangent bundle $(TS^1, g_{S^1})$
 on $Y^{11}$. The lift of the connection $\nabla|_{TS^1}$ to $TS^1$
 is denoted by $\nabla^{S^1}$. 
 Let ${\sf R}$ be an integral power operation $\Z[\Lambda^i, i=0,1,2,\cdots]$
 or $\Z[{\rm Sym}^i, i=0,1,2,\cdots]$, where $\Lambda$ and Sym denote antisymmetric and 
symmetric powers, respectively. 
\footnote{For the Witten genus, we considered only symmetric powers in 
\eqref{virt}.}
The connection $\nabla^{S^1}$ lifts naturally to 
 ${\sf R}(TY^{11})$, which we denote $\nabla^{\sf R}$.  
 For any fiber $S^1$, let $D_{S^1,\sf{R}}$ be the Dirac operator acting on 
 $\Gamma \left( S(S^1)\otimes TS^1 \otimes {\sf R}(TY^{11}|_{S^1})\right)$ defined as
 \(
 D_{S^1, \sf{R}}(\psi_z \otimes {\sf r})=
 c(\xi) \nabla_\xi \psi_z \otimes {\sf r} + c(\xi) \psi_z \otimes \nabla_\xi {\sf r}\;,
 \)
 where $\psi_z\in\Gamma(S(S^1))$ and ${\sf r} \in \Gamma ({\sf{R}}(TY^{11}))$. 
 Note that we are considering the 
 derivative with respect to the 11th direction, something that goes beyond 
 previous treatments.  
 Thus, in a sense, we are considering the vertical components of the 
 Rarita-Schwinger operator, taking into account higher Fourier modes
 'of the circle in the eleventh direction. 
 
 %\paragraph{The kernel of the vertical Dirac operator.} 
\vspace{3mm}
We now consider the zero modes of the operator $ D_{S^1, \sf{R}}$.
Applying \cite{Z1} \cite{Z2}, 
the space $\ker(D_{S^1,\sf{R}})$ is of constant rank and forms a vector bundle over
$X^{10}$ with 
\(
\ker (D_{S^1,\sf{R}})= {\sf R}(TX^{10}\oplus \mathcal{O}_\R)\;.
\) 
Thus, the kernel of the vertical Dirac operator is given by the 
bundle of powers of the tangent bundle. This effect, which improves
what was considered in \cite{MS}, shows that the Fourier modes 
for the M-theory circle correspond to powers of the split tangent 
bundle. 

%\attn{this was no considered in MS}

\paragraph{Geometric representative of the Euler class of the circle bundle.}
 Let $\pi^{S^1}$ be the orthogonal projection on $TS^1$ with respect to 
$g_Y$. In addition to the Levi-Civita connection $\nabla^{LC}$, let 
$\nabla$ be a connection on $TY^{11}$ defined as
(cf. \cite{Bi})
\begin{eqnarray}
\nabla_z v_z&=&\pi^{S^1} (\nabla_z^{LC} v_z)\;,
\qquad \qquad \qquad
\nabla_\mu v_z=\pi^{S^1} (\nabla_\mu^{LC} v_z)\;,
\nonumber\\
\nabla_z v_\mu&=&0\;,
\hspace{3.8cm}
\nabla_\mu v_\nu=\nabla_\mu^X v_\nu\;,
\end{eqnarray}
for $v$ a vector with $z$-component along the eleventh direction or 
$\mu$-component along $X^{10}$. 
As indicated in \cite{DMW-Spinc}, this is more general than the usual 
Kaluza-Klein and Scherk-Schwarz settings for dimensional reduction 
in the sense of allowing $\nabla_\mu v_z$ to be nonzero in general.
Define a tensor $\mathcal{S}$ by 
$
\mathcal{S}=\nabla^{LC} -\nabla
$.
Then, 
for $\xi \in TS^1$ the unit vector field  determined by $g_Y$ and the Spin structure
on $TS^1$,  the formula
$
\mathcal{S}(\xi)(\xi)=0
$ holds \cite{Z2}.
The torsion $\mathcal{T}$ of $\nabla$ defined by 
$
\mathcal{T}_{\mu\nu}=-\mathcal{S}_{\mu\nu} + \mathcal{S}_{\nu \mu}
$.
Let $\xi^* \in T^*S^1$ be the dual of $\xi$. Then, since
$\nabla^{LC}$ is torsion-free, we have the pairing 
\(
\langle \mathcal{T}(U,V), \xi \rangle = d\xi^* (U,V)\;, \qquad U, V \in TX^{10}\;.
\label{Tuv}
\)
Therefore, $\mathcal{T}$ determines a 2-form in $\Lambda^2(T^*X^{10})$ 
such that $\frac{1}{2\pi}\mathcal{T}$ represents the Euler class $e$
of the line bundle corresponding to the circle bundle
 and can be viewed as the Ramond-Ramond 2-form
$F_2$ \cite{DMW-Spinc}.

% Let $\{e_1, \cdots e_{10}\}$ be an orthonormal basis of $TX^{10}$
%and $\{ e^*_1, \cdots, e_{10}^*\}$ be the dual basis, i.e. a basis of 
%$T^*X^{10}$. Set
%\(
%c(\mathcal{T})=\frac{1}{2}\sum_{a,b} e^*_a e^*_b c(T(e_a, e_b))\;.
%\)
%We view this as the $\gamma^{\mu \nu}F_{\mu \nu}$ (in the physicist's 
%component notation) term 
%in the dilatino supersymmetry transformation. We will make more connection 
%below.
%
%

\paragraph{Adiabatic limit of eta invariant for the circle bundle.}
Consider the metric 
$
g_{Y,t}=g_{S^1}\oplus \frac{1}{t}\pi^* g_X
$ on $Y^{11}$
with Levi-Civita connection $\nabla^{L,t}$.
The limit $t\to 0$ of the connection gives (cf. \cite{Bi})
$
\lim_{t\to 0}\nabla^{L,t}=\nabla +\pi^{S^1} \mathcal{S}$.
%where $\mathcal{S}$ is the tensor whose
%antisymmetric part is the torsion $\mathcal{T}$ in (\ref{Tuv}).
Since $\nabla$ preserves the splitting 
$TY^{11}=TS^1\oplus TX^{10}$, then $\pi^{S^1}\mathcal{S}$ does not
contribute to the characteristic forms of  the power bundle
${\sf R}(TY^{11})$.
%\attn{recall this is powers of the bundle}
Let $S_tY^{11}$ be the Spin bundle of $(Y^{11}, g_{Y,t})$ and 
$S_tX^{10}$ the Spin bundle of $(X^{10}, \frac{1}{t}g_X)$. 
Then $S_t Y^{11}=S(S^1)\otimes S_tX^{10}$.
Consider spinors $\Psi \in \Gamma (S_tY^{11})$ 
and polyvectors $v \in \Gamma ({\sf R}_t (TY^{11}))$ on $Y^{11}$.
For any $t>0$, the Dirac operator $D^t_{Y,{\sf R}}$ is the 
differential operator acting on 
$\Gamma (S_t Y^{11} \otimes {\sf R}_t(TY^{11}))$,
\(
D^t_{Y,{\sf R}} (\Psi \otimes v)=
c(\xi) \nabla_\xi ^{L, t} \Psi \otimes v +
c(\xi) \Psi \otimes \nabla_\xi^{L,t} v
+\sqrt{t}\sum_{i=1}^{10}
\left(
c(e_i) \nabla_{e_i}^{L,t} \Psi \otimes v 
+
c(e_i) \Psi \otimes \nabla_{e_i}^{L,t}  v 
\right)\;.
\)
Then $D^t_{Y,{\sf R}}$ is a self-adjoint elliptic first order differential operator on 
$\Gamma (S_t Y^{11} \otimes {\sf R}_t(TY^{11}))$. 
%Let $e$ be the Euler class of the line bundle over $X$. 
The limit $\lim_{t \to 0} \overline{\eta} (D^t_{Y,{\sf R}})$ 
exists in $\R/\Z$. 
The adiabatic limit of the eta invariant 
 is interpreted in \cite{MS} \cite{S-gerbe}
  as the phase of the partition function in the 
  semiclassical limit of the dimensional 
reduction from M-theory to type IIA string theory.  
Assuming the kernel is a vector bundle, as is the case above,
then \cite{Z1}
\(
\lim_{t \to 0}  \overline{\eta} (D^t_{Y,{\sf R}})\equiv
\overline{\eta} (D_{X, {\sf R}(TX\oplus \R)})
+
\frac{1}{(2\pi i)^5}\int_X \widehat{A}(i\mathcal{R}_X)\wedge \widehat{\eta} \quad {\rm mod~}\Z\;,
\label{mod z}
\)
where $\widehat{\eta}$ are the eta-forms on $X^{10}$ 
and $\widehat{A}(i \mathcal{R}_X)$ is the differential form 
representative of the $\widehat{A}$-genus 
defined using the curvature $\mathcal{R}_X$ on $X^{10}$. 
Since $X^{10}$ is even-dimensional, the eta invariant vanishes, and so 
we have
$
\overline{\eta} (D_{X,{\sf R}(TX\oplus \R)}) =
\frac{1}{2} {\rm dim~ker}(D_{X,{\sf R}(TX\oplus \R)})$.
%Given the eta-form above, 
Then, from \cite{Z1} \cite{Z2}, we get
\(
\lim_{t \to 0}  \overline{\eta} (D^t_{Y,{\sf R}})\equiv
\frac{1}{2} {\rm dim~ker}(D_{X,{\sf R}(TX\oplus \R)})
+
\left\langle 
\frac{\widehat{A}(TX^{10}) {\rm ch}({\sf R}(TX^{10}\oplus \R))
\cdot
\tanh (\frac{e}{2}) -\frac{e}{2}
}
{e\tanh \frac{e}{2}}~,~\left[ X^{10}\right]
\right\rangle \quad {\rm mod~}\Z\;.
\label{Zh formula}
\)
This generalizes the discussion in \cite{MS} \cite{S-gerbe} to include
higher modes of the M-theory circle as well as higher powers of the 
bundles.

%\newpage
%%%%%%%%
\subsection{The phase and the modular Atiyah-Patodi-Singer index}
%%%%%%%%

%\subsection{Refinement of the $\eta$-invariant}

%\attn{Some of this has already appeared in ATMP paper}

%\subsubsection{APS for $q$-bundles}
 %\begin{enumerate}
 Our discussion above on higher powers of bundles is related to
 our earlier discussion on $q$-expansions in the context of elliptic genera. 
 We will make use of such a connection in this section, at least formally
 as our main interest is modularity. This complements the discussion in 
 \cite{S-gerbe}.
 
 \vspace{3mm}
 The APS index theorem can be extended purely formally to graded bundles
 or $q$-bundles. Suppose that $E_r$, $r\geq 0$, is a sequence 
 of vector bundles and write
 $
 E_q=\bigoplus_{r\geq 0} E_r q^r
 $, 
 where $q$ is a formal variable. 
 The Dirac operator can be twisted with the finite-dimensional
  $E_r$ to get $D_{E_r}^+=D_r^+$.  
  This can be extended to all of $E_q$ to get
  $
  D_q^+ =\sum_{r\geq 0} D_r^+ q^r ~:~\Gamma (S^+ \otimes E_q) \to \Gamma (S^{-}\otimes E_q)$.  
The index of $D_q^+$ is the formal power series 
 \(
 {\rm ind}(D_q^+)=
 \sum_{r \geq 0} {\rm ind} (D_r^+) q^r=
 \sum_{r \geq 0} (\dim \ker (D_r^+) -\dim {\rm coker} (D_r^+)) q^r\;.
 \)
 The operators $D_r^+$ and $D_q^+$ restricted to the boundary 
 $Y^{11}$ of $Z^{12}$ give the twisted Dirac operators
 $
 D_{r, Y}$ and  $D_{q,Y}=\sum_r D_{r,Y} q^r$.
 The eta invariant of a twisted Dirac operator $D_q$ is the formal series
 $
 \eta_{D_q}(q)=\sum_{r \geq 0} \eta_{D_r} q^r$.
 The $q$-analog of the number of zero modes is 
 \(
 h(q)=\dim \ker (D_q)=\sum_{r\geq 0} \dim \ker (D_r)q^r\;. 
 \)
 % \attn{Combine at lead D and RS as first and second levels in the
% elliptic genus to give mod 2 PF}
 The formal Chern character can also be defined as 
$
{\rm ch}(E_q)=\sum_{r \geq 0} {\rm ch} (E_r) q^r
$.
  Applying the APS index theorem to each twisted Dirac operator
 and adding, keeping track of the grading, gives that the
formal twisted 
Dirac operator $D_q^+$ with the APS boundary 
condition has index (see \cite{Ga})
%cor 3.13
\(
{\rm ind} (D_q^+) =\int_{Z^{12}} {\rm ch}(E_q) \widehat{A}(Z) -
\frac{h_{D_{q,Y}}(q)+ \eta_{D_{q,Y}}(q)}{2}\;.
\)  
Of course the above works for any $4k$-dimensional manifold with 
boundary, with our main cases being $k=3$ for M-theory and 
$k=2$ for the M5-brane.

% \end{enumerate}
 
 \vspace{3mm}
%\paragraph{ Example: Framed manifolds.} 
In \cite{Ga} a new invariant 
 $\eta_W(q)$ of framed manifolds $Y^{4k-1}$ is defined 
 which, as 
 a power series, is also a modular function 
 (modulo the integers).
 For example, there are framings of the spheres 
 $S^{4k-1}=\partial \mathbb{D}^{4k}$ for which $\eta_\cE(q)$ are
 expressed in terms of an Eisenstein series $E_{2k}(\tau)$.

 %\attn{Change of framing is change of string structure?}

   \paragraph{Eta invariants and the Witten genus.}
 
% \begin{enumerate}
 
 Consider The Dirac operator twisted by $S_q TZ^{4k}$. 
% \attn{Why symmetric powers}
 Its index under the 
 global APS boundary condition is related to the Witten genus by 
 \cite{Ga}
 \(
 \phi_W (\partial Z^{4k}) \eta_d (\tau)^{-4k} = \frac{1}{2} (h_W + \eta_W)\;,
 \)
 where $\eta_d(\tau)=\eta_d(q):=q^{1/24}\prod_{n \geq 1}(1-q^n)$ 
 is the Dedekind eta-function and 
 $h_W$ and $\eta_W$ are the $q$-refined number of zero modes and 
 the refined APS eta-invariant, respectively.  
 Since the index and the kernel dimensions $h$ are integers, then
 \(
 \frac{1}{2}\eta_W \equiv \frac{1}{2}\phi_W \eta_d(\tau)^{-4k} \quad {\rm mod}~ \Z\;.
 \)
 
 \paragraph{Example: Disks and spheres.} 
 On $(\mathbb{D}^{4k}, S^{4k-1})$ the eta invariant is 
 $
 \frac{1}{2}\eta_W (S^{4k-1})\equiv E_{2k}(\tau) \eta_d(\tau)^{-4k}$  mod $\Z$. 
 For $k=2$ and $k=3$, respectively, this gives
  \bea
   \frac{1}{2}\eta_W (S^{7})\equiv E_{4}(\tau) \eta_d(\tau)^{-8}  \quad {\rm mod}~ \Z\;,
   \\
 \frac{1}{2}\eta_W (S^{11})\equiv E_{6}(\tau) \eta_d(\tau)^{-12}  \quad {\rm mod}~ \Z\;.
 \eea
 We see that the Eisenstein series $E_4$ and $E_6$ appear in relation to the 
 M5-brane and spacetime in M-theory, respectively. In addition, the former 
 also appears when considering M-theory on decomposable manifolds,
 due to the composite nature of the topological parts of the action.

% \attn{clash of notation $G$ C field}
 
% \end{enumerate}

\vspace{3mm}
We hope to consider in the future further relations between M-theory on one hand and 
modularity and the Witten genus on the other.

%\paragraph{Explicit dependence on the String structure.}
%$S^3 \times X^8$

%%%%%%%%%%
\section{The Results on Global Anomalies and the Partition Functions}
%%%%%%%%%%%
\label{Sec results}
Having provided motivation and corresponding invariants to be used for the 
refinement of anomalies from the Spin to the String case, we now 
start with presenting the results listed in the introduction.

%%%%%%%%%%%%%%%%%%%%%%%%%%
\subsection{Global Anomalies in $D=10$, $N=1$ Supergravity}
%%%%%%%%%%%%%%%%%%%%%%%%%
We consider the following setting. An $E_8\times E_8$ bundle $V_1 \oplus V_2$ on 
a ten-dimensional manifold $M^{10}$. We assume $M^{10}$ to have a 
String structure, ie. that the spin bundle $SM^{10}$ of $M^{10}$ admits
a lifting of the structure group from ${\rm Spin}(10)$ to its 7-connected cover
${\rm String}(10)$.  The condition for such a lift is given by $\lambda(TM^{10})=0$, 
where $\lambda=\frac{1}{2}p_1$ is half the Pontrjagin class. 

\vspace{3mm}
Due to the homotopy type of $E_8$, the $E_8 \times E_8$ bundle on $M^{10}$ 
is completely characterized by the degree four class
\(  
a(V_1 \oplus V_2)= a(V_1) + a(V_2),
\)
where $a=\frac{1}{2} p_1$. 
The anomaly cancelation condition is given by 
\(
\lambda(TM^{10}) - a(V_1 \oplus V_2)=0.
\) 
Assuming $M^{10}$ to admit a String structure implies that 
\(
a(V_1 \oplus V_2)=0.
\)
Note that this does not necessarily imply that each factor is separately 
zero. We will thus take the condition to be
\(
a(V_1) = - a(V_2),
\)
so that $V_1$ has characteristic class $a$ if and only if $V_2$ has 
corresponding class $-a$.  
Note that every class in $K(\Z, 4)$ represents the characteristic class
of an $E_8$ bundle. We have the following fact from \cite{Tools}

\begin{lemma}
The effective action of heterotic string theory is invariant under 
diffeomorphisms $\varphi: M^{10} \to M^{10}$ that admit a lift to the
spin bundle of $M^{10}$ and to $V_1 \oplus V_2$. 
\label{phi}
\end{lemma}

\vspace{3mm}
Global anomalies are concerned with diffeomorphisms and/or 
gauge transformations that are not connected to the identity. The main
question here is the string analog of the question raised in the spin
case in \cite{Tools}: 
{\it Is the effective action of $N=1$ supergravity 
with group $E_8 \times E_8$ invariant under such $\varphi$?}. 
\\
The study of global anomalies requires considering the mapping cylinder
\(
X^{11}= \left(M^{10} \times S^1 \right)_{\varphi}= \left( M^{10} \times I \right)/
(x, 0) \sim (\varphi(x), 1),
\)
and asking whether the bundles extend to $X^{11}$.

\begin{lemma}
{\bf i.}  If $M^{10}$ is a string manifold then so is $X^{11}$.

\noindent {\bf ii.} The bundle $V_1 \oplus V_2$ can be extended to an $E_8 \times E_8$ bundle
over $X^{11}$. 

\noindent {\bf iii.} $X^{11}$ bounds a 12-dimensional string manifold $B^{12}$. 

\end{lemma}

\proof 
The first part follows from the multiplicative behavior of the Spin characteristic classes
under Whitney sum. 
For part (ii), note that  
the String analog of lemma (\ref{phi}) holds. Then the action of $\phi$ on $V_1 \oplus V_2$
leads to the identification of the fiber of $V_1 \oplus V_2$ at $(\phi(x), 1)$ with the fiber
at $(x, 0)$. 
For the third part we have from \cite{Giam} that $M{\rm String}_{11}({\rm pt})=0$ and thus
any 11-dimensional string manifold bounds a 12-dimensional string manifold.
\endproof

We now have

\begin{prop}
The bundle $V_1 \oplus V_2$ extends from $X^{11}$ to an $E_8 \times E_8$ bundle
$W_1 \oplus W_2$ over $B^{12}$. Furthermore, we can have $a(W_1 \oplus W_2)=0$.
\end{prop}

\proof
$V_1$ extends over $B^{12}$ if and only if the cohomology class $a(V_1)$ extends
to a cohomology class in $H^4(B^{12}; \Z)$. Considering the pair $(X^{11}, \beta)$.
This vanishes if $X^{11}$ bounds a String manifold $B^{12}$ over which the class
$\beta \in H^4(X^{11}; \Z)$ can be extended. This means that the pair is an 
element of the cobordism group $\Omega_{11} \left( K(\Z, 4)\right)$. It is shown in
\cite{Hill} that $M{\rm String}_{11}\left( K(\Z, 4)\right)$ is zero. Therefore there is no
obstruction and so $B^{12}$ can always be chosen so that $V_1$ extends to an $E_8$ bundle
$W_1$ over $B^{12}$. 

%\vspace{3mm}
Next we extend $V_2$. Corresponding to the inclusion $\iota :  X^{11} \hookrightarrow
B^{12}$, the tangent bundle of $B^{12}$ decomposes as 
\(
TB^{12}|_{X^{11}}= TX^{11} \oplus \mathcal{N},
\)
where $\mathcal{N}$ is the normal bundle of $X^{11}$ in $B^{12}$. Being a trivial real line
bundle, $\mathcal{N}$ does not change the fact that the class $\lambda$ of $TX^{11}$ is zero, 
i.e.
\( 
\lambda(TX^{11})= \lambda(TX^{11} \oplus \mathcal{{O}})=0.
\)
extends to a trivial cohomology class in $H^4(B^{12}; \Z)$. Since $a(V_1)$ extends
to $a(W_1)\in H^4(B^{12}; \Z)$ then $\beta= -a(W_1)$ is an element of
$H^4(B^{12}; \Z)$. Therefore, $V_1$ and $V_2$ both extend over $B^{12}$. Furthermore,
the extension can be done in such a way that $a(W_1)=-a(W_2)$. 
\endproof

\begin{theorem}
There are no global anomalies for $E_8 \times E_8$ bundles $V_1 \oplus V_2$ 
on string manifolds $M^{10}$.
\end{theorem}

\proof
Given that $(M^{10} \times S^1)_{\varphi}$ bounds a string manifold $B^{12}$ 
over which $V_1 \oplus V_2$ can be extended, the change in the effective action
$S$ under $\varphi$ will be as in the spin case \cite{CMP} \cite{Tools}
\(
\Delta S = 2 \pi i \left[ \int_{B^{12}} \left( tr_1F^2 + tr_2 F^2 - tr R^2 \right) \wedge I_8
- \int_{(M^{10} \times S^1)_{\varphi}} H_3 \wedge I_8
\right],
\label{delta}
\)
where 
\begin{itemize}
\item $F$ is the curvature of the $E_8$ bundle so that $tr_i F^2$ is the 
Chern class of the bundle $W_i$, $i=1, 2$, 
\item $R$ is the curvature of the tangent bundle $TB^{12}$, so that 
$tr R^2$ is the Pontrjagin class,
\item $I_8$ is the Green-Schwarz anomaly
polynomial in characteristic classes of the gauge and tangent bundles, and
hence satisfies $dI_8=0$.
\end{itemize}
Since $V_1 \oplus V_2$ extends to $W_1 \oplus W_2$ on $B^{12}$ such that
$a(W_1) + a(W_2)=0= \lambda(TB^{12})$ and hence (trivially) that 
$a(W_1 \oplus W_2) + \lambda(TB^{12})=0$, then the classes are trivial in 
cohomology and hence exact. Then $H_3$ can be defined to obey
\(
dH_3=tr_1 F^2 + tr_2 F^2 - tr R^2
\label{dh}
\)
not just on the mapping cylinder $(M^{10} \times S^1)_{\varphi}$ but also 
on the bounding manifold $B^{12}$. Inserting (\ref{dh}) in (\ref{delta})
we get
\(
\Delta S= 2 \pi i \left[ \int_{B^{12}} dH_3 \wedge I_8 - 
\int_{(M^{10} \times S^1)_{\varphi}} H_3 \wedge I_8
  \right].
 \)
Now using Stokes' theorem for  $(M^{10} \times S^1)_{\varphi}= \partial B^{12}$
the invariance result $\Delta S=0$ follows.
\endproof

%%%%%%%%%%%
\subsection{Refinement of the M-theory/IIA partition Function} 
%%%%%%%%%%%
The setting for the comparison of M-theory and type IIA string theory is as follows.
M-theory is `defined' on the eleven-dimensional total space $Y^{11}$ of a circle
bundle with base $X^{10}$, a ten-dimensional manifold on which type IIA string 
theory is defined. Both spaces $X^{10}$ and $Y^{11}$ are usually taken to be 
Spin manifolds and they have additional structures on them. On $Y^{11}$ there 
is an $E_8$ bundle $V$. Due to the homotopy type of $E_8$ up to dimension
14, $V$ over $Y^{11}$ is completely characterized by a degree four class $a$
as above.
Various kinds of spinors are defined on both $Y^{11}$ and $X^{10}$. In particular,
for our purposes, there are elements $\lambda$ of $\Gamma(SY^{11} \otimes V)$, 
i.e. sections of 
the spin bundle $SY^{11}$ coupled to the $E_8$ vector bundle. 
On $X^{10}$, in addition to spinors, there are also the Ramond-Ramond (RR) fields
of even degrees. In particular, there is a degree four field $F_4$. Such fields
are images in cohomology of K-theory elements $x$, as they obey the quantization
condition \cite{MW} \cite{FH}
$F:=\sum_{i=0}^5 F_{2i}= \sqrt{\widehat{A}(X^{10})}~ {\rm ch}(x)$. 

\vspace{3mm}
The comparison of the $E_8$ gauge theory description of M-theory to the 
K-theoretic description of type IIA string theory was performed in \cite{DMW}
at the level of the partition functions and is shown to agree upon dimensional
reduction, i.e. integration over the $S^1$ fiber. The comparison involves
those cohomology classes that lift to K-theory and the identification involve
subtle torsion and mod 2 expressions.

\begin{definition}
The {\rm phase} of the M-theory partition function on an eleven-dimensional 
spin manifold $Y^{11}$ is  \cite{DMW} $\Phi_a=(-1)^f(a)$,
where $f(a)$ is the mod 2 index of a Dirac operator coupled to $V$ with characteristic
class $a$.
\end{definition}

\noindent {\bf Remarks}

\noindent {\bf 1.} $f(a)=0$ for $a=0$, in which case $V= \bigoplus^{248} \mathcal{O}$, i.e. 
248 copies of a trivial bundle.

\noindent {\bf 2.} The comparison to type IIA requires a corresponding mod 2 invariant 
in $KO(X^{10})$: For $x \in K(X^{10})$, $j(x)$ is the mod 2 index with values  
in the $KO$ class $x \otimes \overline{x}$. 

\noindent {\bf 3.} The refinement of the partition function to elliptic cohomology $E$ in \cite{KS1}
introduces a mod 2 index with values in the $EO_2(X^{10})=\Z_2[q]$ class 
$x \otimes \overline{x}$, where $x$ is a class in $E(X^{10})$. 

\noindent {\bf 4.} $f(a)$ cannot be a cubic function in $a \in H^4\Z$ since that 
would have dimension 12, which is greater than the dimensions of either
$X^{10}$ or $Y^{11}$.

\vspace{3mm}
The comparison between M-theory on $Y^{11}$ and type IIA string theory on
$X^{10}$ proceeds from the embedding 
\(
SU(5) \times SU(5)/\Z_5 \subset E_8
\label{emb}
\)
so that out of two $SU(5)$ vector bundles $E$ and $E'$ of Chern classes 
$c_2(E)=-a$ and $c_2(E')=-a'$ one builds an $E_8$ bundle whose characteristic
class is $a + a'$. This requires $a$ and $a'$ to be elements of $H^4(X^{10}, \Z)$ 
that lift to K-theory. 
The idea is then to compute $f(a + a')- f(a) - f(a')$. Using the decomposition
(\ref{emb}), this is
\(
\int_{X^{10}} c_2(E) c_3(E')~~{\rm mod}~2= \int_{X^{10}} c_3(E) c_2(E')~~{\rm mod}~2.
\) 
Since $c_1(E)=0$ then $c_3(E)=Sq^2 (c_2(E))$ mod 2, and similarly for $E'$. 
Then the main result on $f(a)$ is that it is a quadratic refinement of a bilinear form, i.e. $f$ satisfies
\cite{DMW}
\(
f(a + a')=f(a) + f(a') + \int_{X^{10}} a \cup Sq^2a'\;.
\label{bilin}
\)
This has an interpretation in terms of cobordism as follows \cite{DMW}.

\vspace{3mm}
\noindent {\bf Remarks}

\noindent {\bf 1.} A result of Atiyah and Singer \cite{AS} states that the mod 2 index of the Dirac operator coupled 
to a vector bundle $V$ on a Spin manifold
$X$ vanishes if $X$ is the boundary of a Spin manifold over which $V$ extends. $f(a)$ can be regarded 
as a $\Z_2$-valued function which vanishes when $a$ extends to a Spin manifold $B^{11}$
bounding $X^{10}$ \cite{DMW}. The class $a \in H^4(X^{10}, \Z)$ can be extended to a bounding manifold
$B^{11}$ if and only if the map $\mu:X^{10} \to K(\Z, 4)$ can be extended to a map 
$\mu':B^{11} \to K(\Z, 4)$, i.e. that $(X^{10}, a)$ is zero as an element of the cobordism 
group $\Omega_{10}^{\rm spin}\left( K(\Z, 4) \right)$. Thus $f(a)$ can be viewed as an 
element of ${\rm Hom} \left( \Omega_{10}^{\rm spin}\left( K(\Z, 4) \right), \Z_2 \right)$.
More precisely, since $f(a)=0$ for $a=0$ then \cite{DMW}
$f \in {\rm Hom} \left( {\widetilde \Omega}_{10}^{\rm spin}\left( K(\Z, 4) \right), \Z_2 \right)$.

\noindent {\bf 2.} A result of Stong \cite{Stong} states that  
${\widetilde \Omega}_{10}^{\rm spin}\left( K(\Z, 4) \right)=\Z_2 \times Z_2$. Thus there
are two invariants: one is linear \cite{Stong}
\(
v(a)= \int_{X^{10}} a \cup w_6= \int_{X^{10}} a \cup Sq^2 w_4,
\)
with $v(a+ a')=v(a) + v(a')$,
and another-- the more relevant one-- is nonlinear \cite{LS}
\(
Q(a + a')= \int_{X^{10}} a \cup Sq^2 a' = \int_{X^{10}} (Sq^2 a) \cup a'. 
\)
$Q(a_1, a_2)$ vanishes if both $a_1$ and $a_2$ can be extended to $B^{11}$ or
if either $a_1$ or $a_2$ is zero. This leads to 
\cite{DMW}: {\it  $Q(a_1 + a_2)$ is a homomorphism from the bordism group
 ${\Omega}_{10}^{\rm spin}\left( K(\Z, 4)\wedge K(\Z, 4) \right)=\Z_2$ to $\Z_2$.
Thus, there is only one cobordism invariant $Q$.}

\vspace{3mm}
The first observation is straightforward
\begin{observation}
There is no DMW anomaly in the M-theory partition function for String manifolds. 
\end{observation}

\proof
The DMW anomaly for Spin manifolds is given by the vanishing of the 
seventh integral Stiefel-Whitney class $W_7=0$ 
\cite{DMW}. Since $W_7= Sq^3 \lambda$ then $\lambda =0$ implies
that $W_7=0$ and hence no anomaly. This fact has also been observed in 
\cite{KS1}.  
\endproof

The invariant in the case of String cobordism is still the Landweber-Stong 
invariant $Q(a_1, a_2)$ \cite{Hill}. The analysis follows that of \cite{DMW}. 
The main result in this section is then  

\begin{theorem}
The phases of M-theory and type IIA string theory coincide not just on Spin manifolds
but also on String manifolds, i.e. $\Phi_M = \Phi_{IIA}$. Consequently, the corresponding
partition functions match.
\end{theorem}

%%%%%%%%%
\subsection{The Cubic Refinement Law}
%%%%%%%%
In \cite{DFM} the phase $\Phi$ of the M-theory partition function 
was interpreted as a cubic refinement of a trilinear form
on the group $\check{H}^4(Y^{11})$ of degree four differential characters
\(
\frac{\Phi_{(123)} \Phi_{(1)} \Phi_{(2)} \Phi_{(3)}}
{ \Phi_{(12)} \Phi_{(13)} \Phi_{(23)} \Phi }
=( \check{a}_1 \check{a}_2 \check{a}_3)(Y^{11}) \in U(1)\;,
\)
where we denoted 
$\Phi_{(ijk)} := 
\Phi([\check{C}] + \check{a}_i + \check{a}_j + 
\check{a}_k)$ and so on.
The validity of this formula requires that $\check{C}$
and all $\check{a}_i$, $i=1,2,3,4$, extend simultaneously 
on the same Spin twelve-dimensional manifold $Z^{12}$. 
In the Spin case, the obstruction to extending 
$(Y^{11}, a_1, \cdots, a_k)$ is
measured by $\Omega_{11}^{\rm spin}(\wedge^k K(\Z, 4))$.
The group is zero for $k=1$ by Stong's result. The result for
$k=2,3,4$ follows from an application of the Atiyah-Hirzebruch 
spectral sequence \cite{DFM}.

\vspace{3mm}
Now we would like to replace the String condition on 
$Y^{11}$ with the String condition, i.e. we will assume that
$\frac{1}{2}p_1(Y^{11})=0$. We know from \cite{Giam} that
$\Omega_{11}^{\rm String}=0$. 
Furthermore, we know from \cite{Hill}
that $\Omega_{11}^{\rm String}(\wedge^k K(\Z, 4))=0$,
for $k=1,2$. Extending the result to the case $k=3, 4$ 
via the Atiyah-Hirzebruch spectral sequence we get

\begin{prop}
The cubic refinement law holds and is defined for String 
eleven-manifolds $Y^{11}$. 
\end{prop}

\vspace{0.5cm}

%%%%%%%%%%%%%%%%%%%%%%%%%%%%%%%%%%%%%%%%%%%%%%%%%%
{\bf \large Acknowledgements}
%%%%%%%%%%%%%%%%%%%%%%%%%%%%%%%%%%%%%%%%%%%%%%%%%%

\vspace{2mm}
\noindent 
The author thanks the American Institute of Mathematics for hospitality and the 
``Algebraic Topology and Physics" SQuaRE program participants 
Matthew Ando, 
John Francis, Nora Ganter, Mike Hill, Mikhail Kapranov, 
Jack Morava, Nitu Kitchloo, 
and Corbett Redden 
for very useful discussions. 
Special thanks are due to Mike Hill for performing the very nice computations in \cite{Hill} and
sharing them during the program. The author would also like to thank the Hausdorff Institute 
for Mathematics (HIM) in Bonn and the organizers of the ``Geometry and Physics" Trimester Program
at HIM for hospitality. 
The first draft of this paper was written in 2008 and the author apologizes for taking so 
long to write the second version. 
The reason is that this version had to wait for some
developments, both by the current author and other authors in order to 
characterize the elliptic invariants needed to replace the classical invariants.

%%%%%%%%%%%%%%%

\end{document}